\documentclass{moriond}
\usepackage{amsmath}
\usepackage{subcaption}

\def\be{\begin{equation}}
\def\ee{\end{equation}}
\def\bea{\begin{eqnarray}}
\def\eea{\end{eqnarray}}

\begin{document}
\title{Pattern-recognition techniques to search for gravitational waves \\ from inspiraling, dark-dressed primordial black holes}

\author{Charchit Kumar Sethi$^{1}$, Andrew L. Miller$^{2,3}$ and Sarah Caudill$^{4}$}
\address{$^{1}$I. Physikalisches Institut, Universität zu Köln, Köln, Germany\\
$^{2}$Nikhef -- National Institute for Subatomic Physics,
Science Park 105, 1098 XG Amsterdam, The Netherlands \\
$^{3}$Institute for Gravitational and Subatomic Physics (GRASP),
Utrecht University, Princetonplein 1, 3584 CC Utrecht, The Netherlands\\
$^{4}$Department of Physics, University of Massachusetts, Dartmouth, Massachusetts 02747, USA}
\maketitle\abstracts{Primordial black holes (PBHs) are compelling dark matter (DM) candidates, but current constraints suggest they cannot compose all of DM. This implies that additional DM components could coexist with PBHs, one of which  could form “dark dresses” (DDs) around PBHs. DDs would cause PBH binaries to experience dynamical friction (DF), which would accelerate their inspirals with respect to those in vacuum. Ignoring DF effects in matched-filtering searches could lead to significant sensitivity loss, especially in systems with asymmetric mass-ratios of $q \sim 10^{-3}$. 
We thus show that a method designed to find time-frequency power-law tracks from inspiraling PBHs in vacuum could actually handle the presence of DDs with minimal modifications. This method, the generalized frequency-Hough (GFH), maps points in the detector's time-frequency plane to lines in the source parameter space. We show that this pattern-recognition technique can recover simulated DD signals in Gaussian noise, marking an important step forward in developing DM-aware methods beyond matched filtering.}

\section{Introduction} The nature of DM remains one of the key puzzles in cosmology. PBHs, formed from early-universe density fluctuations, have been proposed as DM candidates. Though observational constraints limit their contribution across most of the mass spectrum\cite{Green_2021}, they may still represent a fraction of DM. In multi-component models, PBHs could coexist with other forms of DM that form gravitationally bound “dark dresses” (DDs) around them\cite{Brito_2020}. These DDs could modify the inspiral of compact objects via DF, leading to phase evolutions in gravitational-wave (GW) signals that differ from the vacuum case. For asymmetric mass ratio systems ($q \sim 10^{-3}$), this effect could cause up to 70\%  sensitivity loss if not accounted for in standard matched-filtering analyses\cite{PhysRevD.107.083006}. To address this, we simulate DD-PBH inspirals and assess whether the GFH can recover these signals, and demonstrate that DM effects can be incorporated in GW searches.

\section{GWs from Dark-Dressed Inspirals} In the presence of a DD, inspiraling compact objects experience DF in addition to GW emission. This leads to a modified spin-up equation (the rate of change of the frequency):
\begin{equation}
\dot{f}_{\rm DD} = C_{\rm GW} f^{11/3} + C_{\rm DF} f^{3/2} \equiv \dot{f}_{\rm Vac} + \dot{f}_{\rm DF},
\label{eq:spinup}
\end{equation}
where $\dot{f}_{\rm Vac}$ corresponds to vacuum GW-driven evolution and $\dot{f}_{\rm DF}$ arises from DF due to the DD halo. $C_{\rm GW} \equiv \frac{96}{5} \pi^{8/3} \left(\frac{G\mathcal{M}}{c^3}\right)^{5/3}$ depends on the chirp mass $\mathcal{M}$, and $C_{\rm DF} \propto q\,\rho_6\,r_6^{9/4}\,M^{-3/4} \log\Lambda\,\xi$, where $q$ is the mass ratio, $\rho_6$ the DM density in $M_\odot/{\rm pc}^3$, $r_6$ the radius in units of ${\rm pc}$, $\log\Lambda$ the Coulomb logarithm, and $\xi$ the DM fraction.
\begin{figure}[ht!]
    \centering
    \includegraphics[width=0.6\linewidth]{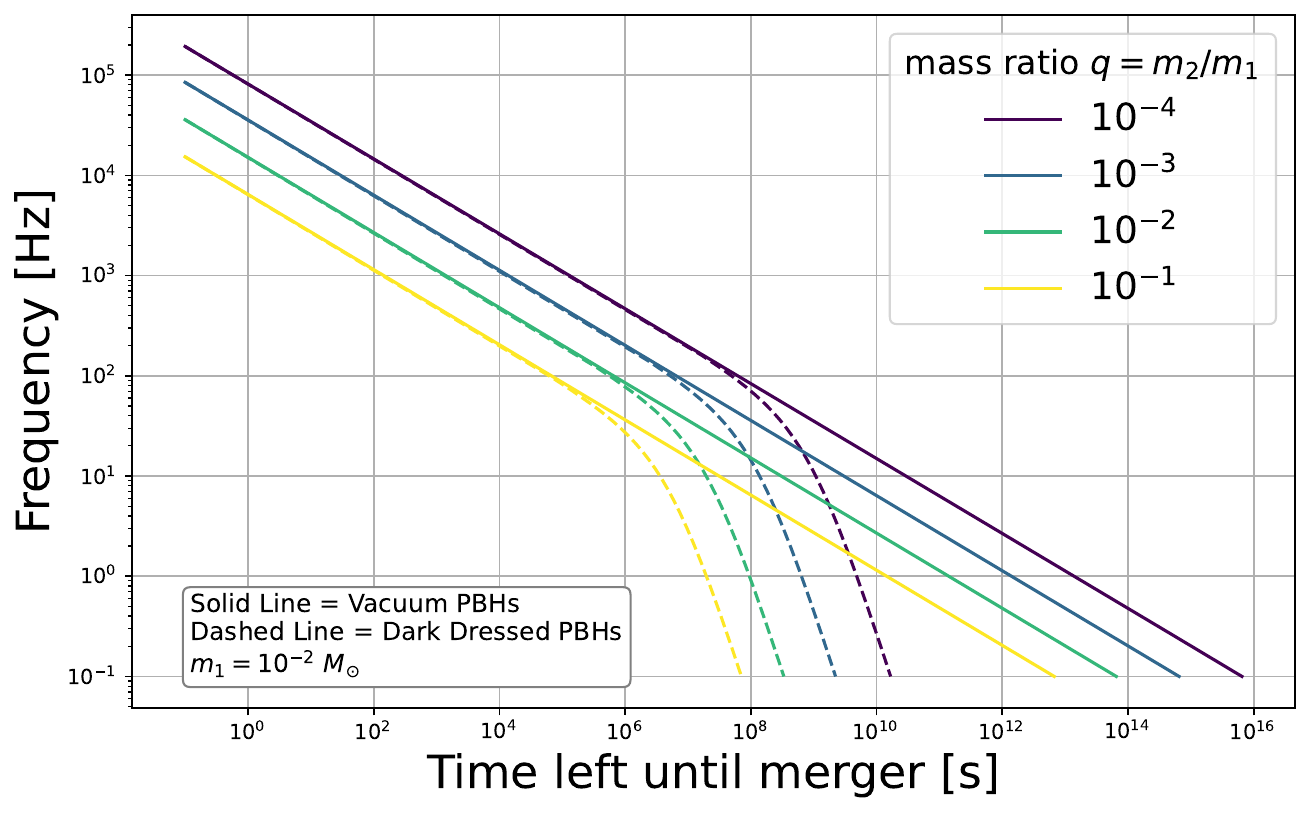}
    \caption{Comparison of time-frequency evolution PBH binaries in vaccuum and with DDs, for different mass ratios. The impact of DDs is visible at low frequencies (below 100 Hz).}
    \label{fig:enter-label}
\end{figure}
\section{Method} We modeled the inspiral of DD-PBH binaries by including both GW emission and DF from surrounding DM. This results in a frequency evolution that deviates from the standard vacuum power law. Simulated GW signals were generated across various mass ratios $(q)$ and then injected into Gaussian noise. We analyzed the data using the GFH, which maps time-frequency tracks to source parameters assuming a particular power-law evolution\cite{MILLER2021100836}. To account for regimes where DF and vacuum effects are comparable, we introduced an effective constant $C_{\rm eff}\equiv\frac{\dot{f}_{\rm dd}}{f^n}$ to approximate Eq.~\eqref{eq:spinup} as a single power law, capturing both $C_{\rm GW}\, \&\, C_{\rm DF}$ contributions and enabling robust recovery of GW signals even in mixed-dynamics scenarios.
\section{Results} 
\begin{figure}[ht!]
\begin{minipage}{0.2\linewidth}
\centerline{\includegraphics[width=1.5\linewidth]{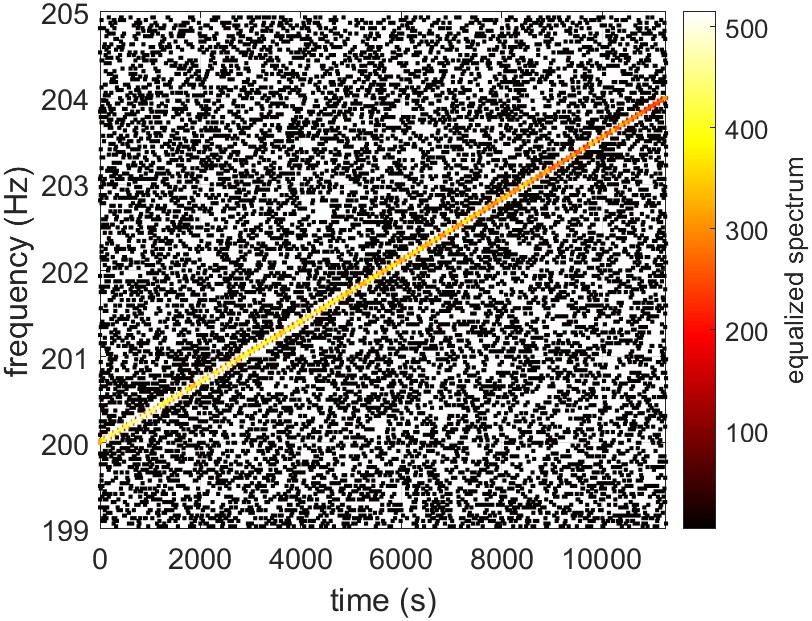}}
\vspace{1mm}
\centerline{\textbf{(a)} Peakmap}
\end{minipage}
\hfill
\begin{minipage}{0.2\linewidth}
\centerline{\includegraphics[width=1.5\linewidth]{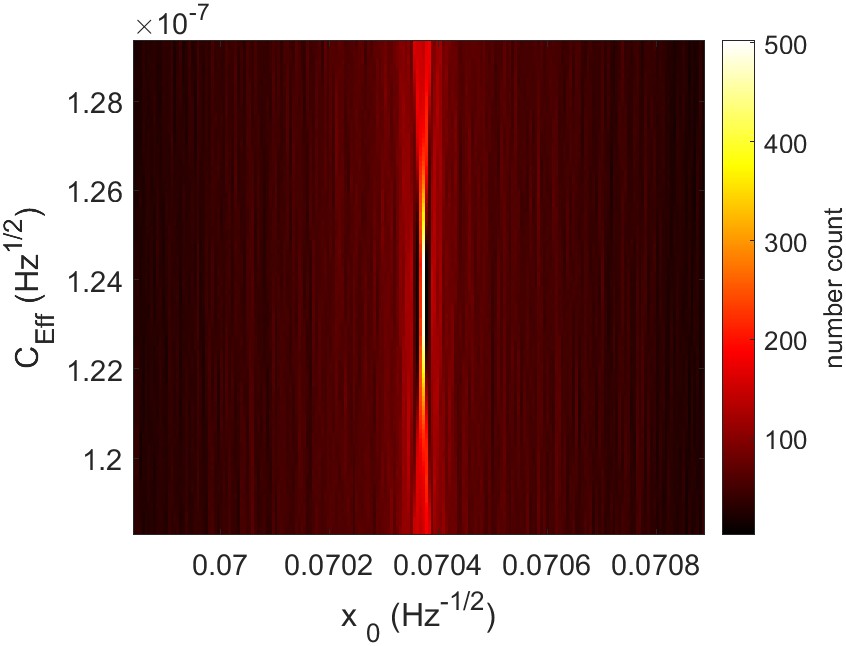}}
\vspace{1mm}
\centerline{\textbf{(b)} Hough Map}
\end{minipage}
\hfill
\begin{minipage}{0.2\linewidth}
\centerline{\includegraphics[width=1.5\linewidth]{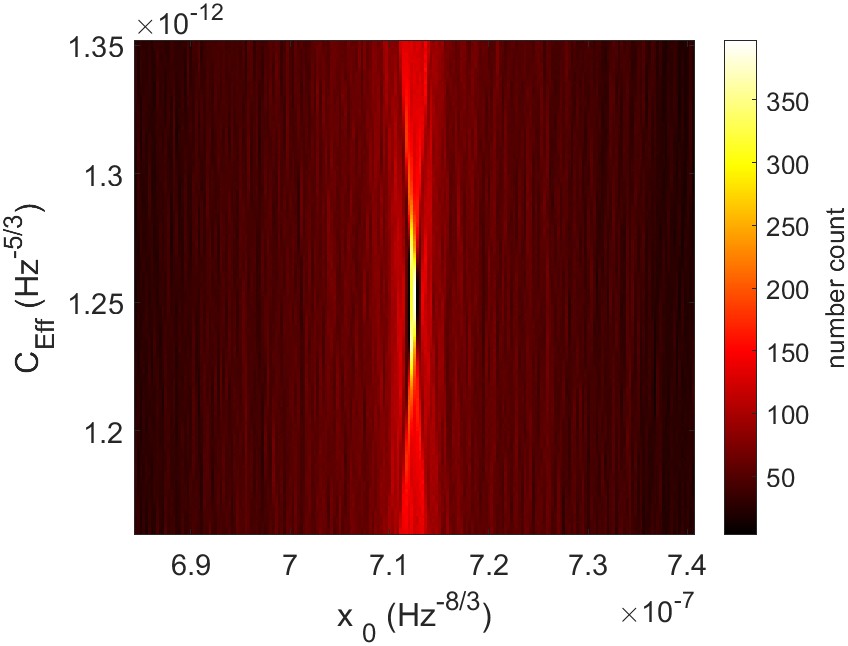}}
\vspace{1mm}
\centerline{\textbf{(c)} Hough Map}
\end{minipage}
\caption[]{Signal from DD PBHs inspiral with mass ratio $q = 10^{-1}$ injected into Gaussian noise in the regime where both the DF and vacuum GW emission are comparable.  The two Hough maps, for $n=11/3$ and $n=3/2$, show successful recovery of the parameters of the injected signal.}
\label{fig:GFHT recovery}
\end{figure}

We injected numerically simulated DD-PBH inspirals into Gaussian noise in the regime where both effects are comparable. Using a new way of constructing the grid in the parameter space and an effective constant $C_{\rm Eff}$, the GFH was able to reliably detect DD-PBH signals -- see Fig. \ref{fig:GFHT recovery}. 
\section{Conclusions} Our method successfully identifies DD-PBH candidates across all inspiral regimes—vacuum-dominated, DF-dominated, and mixed regimes (where vacuum and DF effects are comparable)—demonstrating its potential for detecting complex gravitational wave signatures altered by DM interactions. In the future, we will do an injection campaign in synthetic Einstein Telescope data to determine the efficiency of the GFH in comparison to matched filtering. We will also consider the impact of other forms of DM on the PBH inspiral.
\small
\section*{Acknowledgment} The authors gratefully acknowledge the support and guidance of Philippa S. Cole and Bradley J. Kavanagh throughout the course of this research. Their expertise, constructive feedback, and thoughtful discussions significantly contributed to the development and refinement of this work. C.K.S. also thanks Nikhef for providing funding support, without which participation in this conference would not have been possible.
\begingroup
\footnotesize
\section*{References}

\bibliography{moriond}

\begin{thebibliography}{1}

\bibitem{Green_2021}
Anne~M Green and Bradley~J Kavanagh.
\newblock Primordial black holes as a dark matter candidate.
\newblock {\em Journal of Physics G: Nuclear and Particle Physics}, 48(4):043001, February 2021.

\bibitem{Brito_2020}
Richard Brito, Vitor Cardoso, and Paolo Pani.
\newblock {\em Superradiance: New Frontiers in Black Hole Physics}.
\newblock Springer International Publishing, 2020.

\bibitem{PhysRevD.107.083006}
Philippa~S. Cole, Adam Coogan, Bradley~J. Kavanagh, and Gianfranco Bertone.
\newblock Measuring dark matter spikes around primordial black holes with einstein telescope and cosmic explorer.
\newblock {\em Phys. Rev. D}, 107:083006, Apr 2023.

\bibitem{MILLER2021100836}
Andrew~L. Miller, Sébastien Clesse, Federico {De Lillo}, Giacomo Bruno, Antoine Depasse, and Andres Tanasijczuk.
\newblock Probing planetary-mass primordial black holes with continuous gravitational waves.
\newblock {\em Physics of the Dark Universe}, 32:100836, 2021.

\end{thebibliography}
\endgroup

\end{document}